# Phase decomposition in an Fe-Cr alloy at 402°C: Mössbauer spectroscopic study

S. M. Dubiel[1*] and J. Żukrowski[2]

[1]AGH University of Science and Technology, Faculty of Physics and Applied Computer Science, al. A. Mickiewicza 30, 30-059 Kraków, Poland, [2]AGH University of Science and Technology, Academic Centre of Materials and Nanotechnology, al. A. Mickiewicza 30, 30-059 Kraków, Poland,

**Abstract**

Phase decomposition in strained and strain-free samples of $Fe_{84.85}Cr_{15.15}$ induced by an isothermal vacuum annealing at 402°C up to 3888 h was studied by means of the Mössbauer spectroscopy. The measured $^{57}$Fe-site spectra were analyzed with two independent procedures viz. a hyperfine field distribution and a two-shell superposition methods. Information relevant to the kinetics of the decomposition process, concentration of Cr in the Fe-rich phase, short-range ordering and magnetic texture was obtained based on values of spectral parameters. In particular the following quantities pertinent to the kinetics were found: the activation energy, equal to 1666 kJ/mol for the strained and to 1388 kJ/mol for the strain-free samples, the Avrami exponent equal to 2.5 for both samples, the rate constant equal to 0.0011 and 0.0028 h$^{-1}$ for the strained and the strain-free samples, respectively. The concentration of Cr in the Fe-rich phase was found to be 11.6(4) at. % for the strained sample and 13.2(4) at.% for the strain-free one. Significant differences in the short-range parameters were revealed for the two samples. An average angle between the magnetization vector and the normal to the sample's surface was equal to 52.7° for the strained sample and to 57.5° for the strain-free one.

* Corresponding author: **Stanislaw.Dubiel@fis.agh.edu.pl**

## 1. Introduction

Phase decomposition into Fe-rich ($\alpha$) and Cr-rich ($\alpha$') phases is one of the two main reasons (the other one being a sigma-phase precipitation) for the so-called *475$^o$C embrittlement* that may occur in technologically important structural materials such as ferritic stainless steels based on Fe-Cr alloys, if subjected to temperature between 280 and 500 $^o$C [1]. This frequently occurs in practice, as different devices manufactured from such materials (e. g. heat exchangers, gas turbines, etc.) work at service at elevated temperatures. The precipitation of $\alpha$' leads to a progressive deterioration of materials' useful properties such as hardening or reduction of their toughness. In the phase-diagram of the Fe-Cr system, a zone where the decomposition takes place is designated as *miscibility gap*. The knowledge of its borders, and, in particular, its Fe-rich line, is of a basic importance for predicting and understanding of the steels behaviour at elevated temperatures. Theoretically calculated phase diagrams of the Fe-Cr alloy system are at variance with each other, in particular, as far as the Fe-rich border of the miscibility gap is concerned [2-5]. A validation of these prediction is hardly possible as no experimental data are available below ~415 $^o$C and those above this temperature widely scatter making the verification unable [4]. The practical reason for the lack of experimental data in the low temperature range (< ~415 $^o$C) lies likely in a low rate of the decomposition process that drastically decreases with temperature. For example the rate constant at 415$^o$C is 7-fold smaller than the one at 450$^o$C [6]. In these circumstances relevant measurements aimed at determination of the Cr solubility limit at temperatures lower than 415$^o$C are timely and justified. Mössbauer spectroscopy (MS) is one of relevant methods that can provide proper information on the issue [6,7]. Analysis of Mössbauer spectra can give not only data on the concentration of Cr in the Fe-rich phase, hence the solubility limit, but it also permits to extract information pertinent to a kinetics of the phase decomposition, local concentration of Cr and short-range ordering [6,7,8,9].

This paper presents information on the phase decomposition in an $Fe_{84.85}Cr_{15.15}$ model alloy as obtained by means the Mössbauer spectroscopic study. Two samples were investigated viz. one in a strained and the other in a stran-free state and the decomposition process was activated by isothermal annealing at 402$^o$C.

## 2. Samples

Two samples of a model EFDA alloy of Fe-Cr14 (chromium content 15.15 at%) were used in form of ~20 x 20 mm$^2$ foil with a thickness of ~30 μm. The foils were obtained by cold-rolling (CR) of ~1 mm thick slabs cut from original ingots. One sample, called hereafter "strained", was studied as received after the CR treatment, while another sample, called "strain-free", was vacuum annealed at 700°C for 2 h followed by a slow cooling in furnace. Both samples were next isothermally annealed in a vacuum (~10⁻⁵ mb) at 402°C. The process of the phase decomposition at the chosen temperature was expected to be very slow (it took ~20 hours at 415°C to complete the process [6]), so the Mössbauer spectra in the present case were measured on the samples annealed in a 28-days cycle. The final measurement was done on the samples annealed for ~162 days (3888 h).

## 3. Spectra analysis

$^{57}$Fe Mössbauer spectra, of which some examples are shown in Figs. 1 and 2, were measured at room temperature in a transmission geometry. The measured spectra were analyzed in terms of two methods: (I) Hyperfine magnetic field distribution approach described in detail elsewhere [10], and (II) Two-shell superposition method outlined in detail elsewhere [7,11]. Concerning (I), a linear relationship between the hyperfine field and the isomer shift was assumed following the experimental finding [10]. The fitting procedure yielded hyperfine magnetic field distribution curves, examples of which can be seen in Figures 1 and 2 (right-hand panels). Their integration gave an average hyperfine field, $\langle B \rangle = \int p(B)dB$, the quantity pertinent to determination of the average Cr content in Fe-Cr [7]. The fitting procedure II enabled an insight into a distribution of Cr atoms within the first-two neighbor shells around the probe Fe atoms. This chance follows from the fact that the Fe-site hyperfine field, $B$, is not only sensitive to Cr atoms present in the first (*1NN*) and in the second (*2NN*) shells but even a clear distinction between Cr atoms situated in *1NN* and those in *2NN* can be made [5]. Thanks to this property, the analysis of the Mössbauer spectra in terms of method (II) yielded an information on probabilities of various atomic configurations, *P(m,n)*, *m* being a number of Cr atoms in *1NN*, and *n* that in *2NN*. Knowing the atomic configurations, *(m,n)* and the *P(m,n)*-values enables, in turn, determination of an average number of Cr atoms in *1NN*, *<m>=Σm·P(m,n)*, that in *2NN*, *<n>=Σn·P(m,n)*, as well as in both shells, *<m+n> = <m>+<n>*. In turn, the knowledge of *<m>, <n>* and *<m+n>* found in that way combined with the corresponding quantities expected for the random distribution, *<m>ᵣ=8x, <n>ᵣ=6x,* and *<m+n>ᵣ=14x*, permits determination of short-range (SRO) parameters using the following formulae:

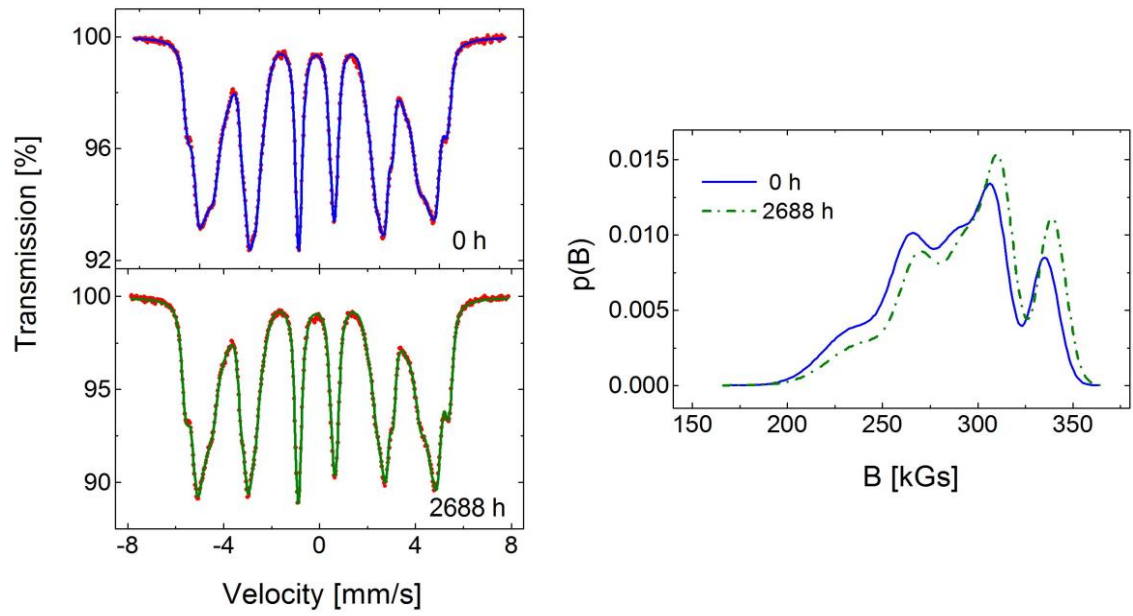

**Fig. 1** $^{57}$Fe spectra recorded at RT on non-annealed (0 h) and annealed (2688 h) strained sample, and corresponding hyperfine field distribution curves.

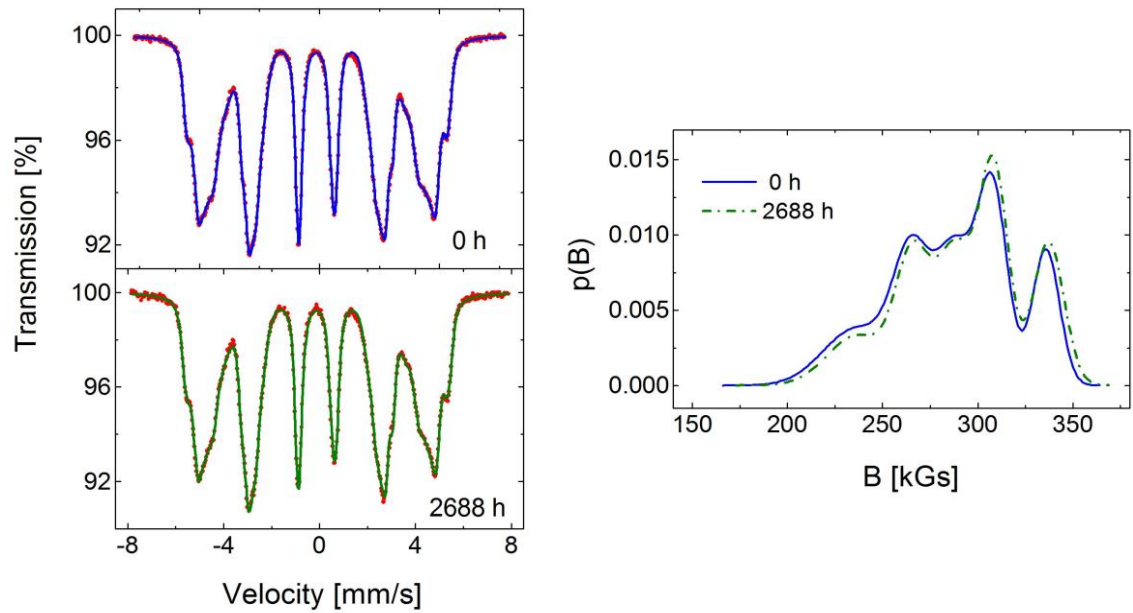

**Fig. 2** $^{57}$Fe spectra recorded at RT on non-annealed (0 h) and annealed (2688 h) strain-free sample, and corresponding hyperfine field distribution curves.

$$\alpha_1 = 1 - \frac{\langle m \rangle}{\langle m_r \rangle} \quad (1a)$$

$$\alpha_2 = 1 - \frac{\langle n \rangle}{\langle n_r \rangle} \quad (1b)$$

$$\alpha_{12} = 1 - \frac{\langle m + n \rangle}{\langle m_r + n_r \rangle} \quad (1c)$$

The spectra were analyzed assuming the effect of the presence of Cr atoms in the *1NN-2NN* vicinity of the $^{57}$Fe probe nuclei on the hyperfine field, *B*, and the isomer shift, *IS*, was additive i.e. $X(m,n) = X(0,0) + m\Delta X_1 + n\Delta X_2$, where *X=B or IS*, $\Delta X_k$ is a change of *X* due to one Cr atom situated in *1NN* (*k*=1) and in *2NN* (*k*=2). The total number of possible atomic configurations *(m,n)* is equal to 63, but for *x* = 15.15 at% most of them have vanishingly small probabilities, so 17 most probable (according to the binomial distribution) were selected to be included into the fitting procedure (their overall probability was > 0.99). However, their probabilities (related to spectral areas of sextets corresponding to the chosen configurations) were treated as free parameters. Free parameters were also *X(0,0)*, line widths and their relative ratios. On the other hand, fixed were values of $\Delta X_k$ 's viz. $\Delta B_1$= -30.5 kOe, $\Delta B_2$= -20.5 kOe, $\Delta IS_1$ = -0.02 mm/s, and $\Delta IS_1$= -0.01 mm/s [7,8,9,11].

## 4. Results and discussion

The best-fit spectral parameters obtained by analyzing the spectra with the procedures described in section 3 are displayed in Table 1. The table contain also other quantities derived from these parameters that are discussed below. In particular, in should be noticed that the corresponding values of the average hyperfine field obtained with the two fitting procedures, $<B>_I$ and $<B>_{II}$ agree well with each other, a feature that can be regarded as a proof that the much more complex but also much more detailed analysis of the spectra with the procedure I yielded correct values of spectral parameters.

**Table 1**
Best-fit spectral parameters as obtained for strained and strain-free samples annealed at 402$^o$ C for various time, *t*: probability, P (0,0), isomer shift, IS(0,0) and the hyperfine field, H(0,0) of the atomic configuration with no Cr atoms within the first-two neighbor shells; the average isomer shift, <IS>, the average hyperfine fields, $<B>_{I,II}$, and the number of Cr atoms in 1NN,

<m>, and in 2NN, <n>, neighbor-shells; the ratio between the amplitude of the second (fifth) and the third (fourth) lines in the sextet, C2/C3, and the average angle between the magnetization vector and the gamma rays, θ.

| t [h] | P(0,0) | IS(0,0) [mm/s] | <IS> [mm/s] | B(0,0) [kGs] | <B>$_I$ [kGs] | <m> | <n> | <B>$_{II}$ [kGs] | C2/C3 | θ [°] |
|---|---|---|---|---|---|---|---|---|---|---|
| | | | | Strain-free | | | | | | |
| 0 | 0.1327 | -.0986 | -.129 | 336.5 | 285.6 | 1.0459 | .9733 | 286.5 | 2.1863 | 57.3 |
| 672 | 0.1416 | -.1006 | -.131 | 340.3 | 290.4 | 1.0307 | .9494 | 289.8 | 2.1546 | 56.7 |
| 1344 | 0.1437 | -.0995 | -.130 | 340.1 | 290.4 | 1.0966 | .8310 | 290.9 | 2.2074 | 57.5 |
| 2016 | 0.1448 | -.0997 | -.130 | 340.4 | 290.9 | 1.1093 | .8489 | 292.1 | 2.2083 | 57.5 |
| 2688 | 0.1524 | -.0999 | -.130 | 339.3 | 289.8 | 1.0687 | .8315 | 292.2 | 2.2232 | 57.7 |
| 3888 | 0.1486 | -.0985 | -.128 | 340.5 | 290.9 | 1.0689 | .8495 | 291.4 | 2.2280 | 57.8 |
| | | | | Strained | | | | | | |
| 0 | 0.1275 | -.0990 | -.129 | 335.6 | 284.9 | .9112 | 1.1779 | 285.4 | 2.0861 | 52.9 |
| 672 | 0.1536 | -.0990 | -.129 | 335.6 | 284.9 | .9112 | 1.1779 | 285.4 | 2.0861 | 52.9 |
| 1344 | 0.1637 | -.1000 | -.129 | 340.6 | 293.0 | .9570 | .9487 | 293.8 | 1.8254 | 52.3 |
| 2016 | 0.1681 | -.0994 | -.127 | 340.2 | 294.1 | .8518 | 1.0315 | 294.4 | 1.8495 | 52.7 |
| 2688 | 0.1720 | -.0983 | -.126 | 340.4 | 294.8 | .8846 | .9776 | 295.2 | 1.8827 | 53.1 |
| 3888 | 0.1657 | -.0984 | -.126 | 339.6 | 294.4 | .9179 | .8805 | 294.8 | 1.8908 | 53.2 |
| 3888 | 0.1665 | -.0986 | -.126 | 340.6 | 295.0 | .8971 | .9344 | 294.8 | 1.7946 | 52.6 |

### 4.1. Kinetics of phase decomposition

A dependence of $<B>=(<B>_I+<B>_{II})/2$ on the annealing time, $t$, for the strained and the strain-free samples are presented in Fig. 3, can be used to follow a kinetics of the phase decomposition. A characteristic saturation-like behavior can be regarded as indication of the termination of the process. It could be well fitted with the following Johnson-Mehl-Avrami-Kolmogorov (JMAK) equation:

$$<B>=<B_o>+b[1-\exp(-kt)^n] \qquad (2)$$

Where $<B_o>$ is the value of the average hyperfine field for the non-annealed sample, $k$ is the rate constant, $n$ is the Avrami exponent, and $b$ is a free parameter.

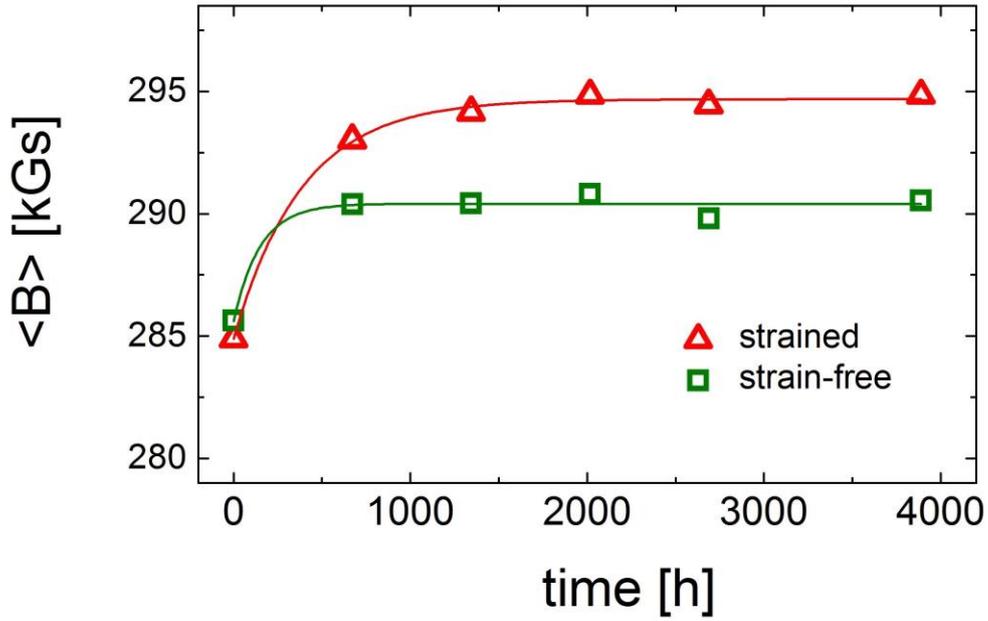

**Fig. 3** Dependence of the average hyperfine field, <*B*>, on the annealing time, *t*, for the studied samples annealed at 402 °C. The best fit to the data in terms of equation (2) is indicated by solid lines.

The best-fit parameters pertinent to eq. (1) and the data shown in Fig. 3 are displayed in Table 2.

Table 2 The best-fit parameters obtained by amalyzing the annealing time dependence of the average hyperfine field, <B>, in terms of eq. (1).

| Sample | $B_o$ [kOe] | b | k [$h^{-1}$] | n | E [kJ/mol] |
|---|---|---|---|---|---|
| Strained | 284.9(1) | 9.8(1) | 0.0011(1) | 2.6(2) | 1666 |
| Strain-free | 285.6(3) | 4.8(4) | 0.0028(1) | 2.4(2) | 1388 |

An obvious difference in the kinetics of the phase decomposition evidenced visually in Fig. 3 is reflected in the rate constant, *k*, which for the strained sample is almost threefold smaller. Yet much greater difference in the rate constant shows up if one compares its values determined on the same alloy i.e. having the same composition and being in the same strained state but measured at different temperatures. For example, at 450°C *k* =2.2 $h^{-1}$ was found [6]. This means that a merely ~10% decrease of temperature causes $2·10^3$ – fold decrease in the rate constant. The knowledge of the rate constant measured at two different temperatures, $T_1$ and $T_2$, enables, via the Arrhenius equation, determination of the activation energy, $E$, using the following formula:

$$E = \frac{T_1 T_2}{T_2 - T_1} k_B \ln(\frac{k_2}{k_1}) \qquad (2)$$

where $k_B$ stays for the Boltzmann constant.

To determine $E$ the value of $k_2$=0.3 as found previously at $T_2$=415°C [6]. In this way we arrived at the $E$-values given in Table 1. They are significantly different between themselves, which means that the metallurgical state of the sample is of importance as far as the kinetics of the phase decomposition is concerned. Noteworthy, the presently found values of $E$ are 6-7 times higher than the activation energy determined previously for $T_1$=415°C and $T_2$ = 450°C [6] (by the way, the correct value in the latter case should read 235.4 kJ/mol, instead of 122 kJ/mol as erroneously given in [6]). These figures clearly show how dramatically a decrease of $T$ slows-down the process of the phase decomposition.

The value of $n$ (Avrami exponent) is usually used to provide information on mechanism(s) responsible for the transformation. Its values obtained in this study for both samples are equal to 2.5. This means that the mechanism driving the phase decomposition in the studied samples has the same character viz. mixed nucleation and volume diffusion controlled growth [12].

### 4.2. Concentration of Cr in the Fe-rich phase

The value of $<B>$ in saturation, $<B_S>$, hence after the decomposition process has practically terminated, can be used for determination of the content of Cr in the Fe-rich phase. The latter can be regarded as the solubility limit at a given temperature of annealing. For this purpose a relationship between $<B>$ and the chromium content, $x$, which is monotonous and smooth [6] can be used. According to this relationship, $<B>$ decreases quasi-linearly with $x$, thus the knowledge of $<B>$ associated with the Fe-rich phase (designated here as $<B_S>$) permits calculation of the Cr content in that phase. Another feature of the $p(B)$- curves that signifies the phase decomposition is an increase of the intensity of the peak situated at ~33 T which can be associated with those Fe atoms that have no Cr atoms in their neighborhood (in our notation (0,0) atomic configuration). The increase of its intensity with the annealing time testifies to a decrease of the chromium concentration in the Fe-rich phase. Another feature in the $p(B)$- curves that shows up on annealing is a shift of the distribution curve towards a higher values of $B$. This aspect is especially well seen in the strained sample – Fig. 1(low panel). It means an increase of the average value of $B$, $<B>$, hence a decrease of the Cr content. In the case of the strained sample the annealing-induced increase of <*B*> is equal to 10 kOe while for the strain-free sample it is equal to 5 kOe. The former corresponds to a reduction of the Cr content by 3.6 at%, while the latter to 1.8 at%. Taking into account that the average Cr concentration in the studied samples is 15.15 at.%, one arrives at 11.55 at.% for the Cr content in the Fe-rich phase and at 13.35 at.% for the stain-free one. These figures can be regarded as the solubility limit for Cr in iron at $402^oC$. For comparison, we found 10.5 (5) at.% at $415^oC$ and 13.3(4) at.% at $450^oC$, in both cases for the strained samples [6].

### 4.3. Short range order

The analysis of the spectra in terms of the procedure (II) and the use of equations (1a)-(1c) enabled determination of the SRO-parameters. They are presented vs. annealing time in Fig. 4.

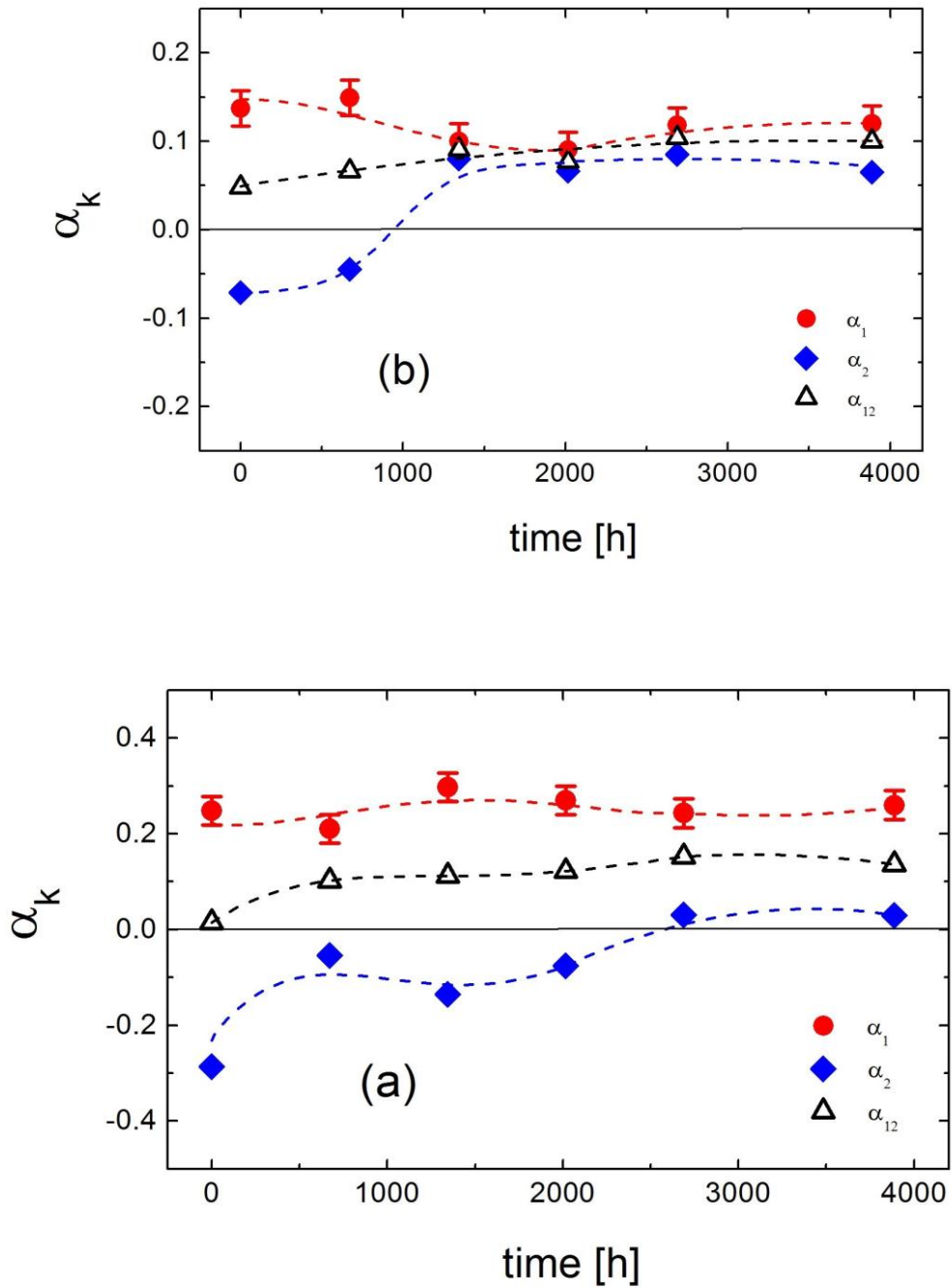

Fig. 4 SRO parameters for the strained (a) and the strain-free (b) samples versus the annealing time. The lines are the guide to the eyes.

It is evident that the SRO-parameters are characteristic of a given neighbor shell, and for the given shell they also significantly depend on the sample. Concerning the strained sample, $\alpha_1 > 0$, and its value hardly changes with the annealing time. This means that the population of Cr atoms in the 1NN-shell is higher than expected for the random distribution and the annealing time does not practically affect this population. At variance with this behaves $\alpha_2$, namely, it is negative and its amplitude decreases with the annealing time reaching eventually zero. This implies that the Cr atoms preferentially occupy the 2NN-shell, but upon annealing their number decreases reaching finally the value characteristic of the random distribution. The SRO-parameter $\alpha_{12}$, which gives information on the distribution of Cr atoms within the two neighbor shells, is close to zero at the start but on annealing it progressively increases reaching eventually saturation – see Fig.5. This indicates a gradual decomposition process leading to a clustering of Cr atoms. From the behaviors of $\alpha_1$ and $\alpha_2$ is follow that predominantly Cr atoms present in the 2NN shells take part in the decomposition process and form the Cr-rich phase ($\alpha$'). This indicates a weaker bonding between Fe and Cr atoms being the second-nearest neighbors and it sounds plausible.

Regarding the strain-free sample, there are some similarities and some differences in comparison to the strained sample. Touching the former, $\alpha_1$ is also positive but its values are lower and show some decrease with the annealing time, $t$, between 28 and 56 days and for longer annealing times it saturates. The $\alpha_{12}$ exhibits analogous behavior like the one in the strained sample viz. it is positive and grows with $t$, yet it values are systematically lower i.e. the degree of clustering in this sample is lower. The different behavior demonstrates $\alpha_2$ as it shows a crossover from a negative to a positive value between 28 and 56 days of annealing, and remains constant for longer annealing times. This means that the population of Cr atoms in the 2NN shell in this sample significantly decreased within the above given period of annealing, and stabilized at a level lower than the one expected for the random distribution. The data presented in Fig. 4 clearly demonstrate that the metallurgical state of a Fe-Cr alloy has to be taken into account in order to properly understand mechanism of the phase decomposition. To highlight this aspect a comparison between $\alpha_{12}$ obtained for the strained and the strain-free sample is displayed in Fig. 5. The difference exists already at the start viz. the degree of the clustering is higher in the strain-free sample.

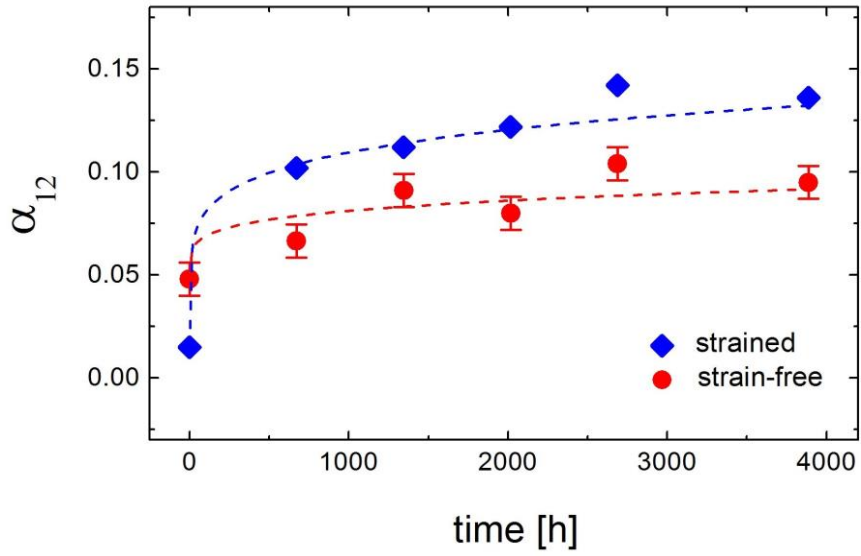

Fig. 5 Short-range parameter, $\alpha_{12}$, versus annealing time for the studied samples. The lines represent the best fits in terms of the JMAK-like equation.

**4.4. Local Cr concentration**

Annealing-induced changes in the distribution of Cr atoms can be also expressed in terms of a local concentration of Cr. For this purpose one can use values of *P(0,0)* and, assuming that the distribution of Cr atoms is random, one can derive therefrom the local Cr content, $x(0,0) = 1 - \sqrt[14]{P(0,0)}$. Such-obtained values of *x(0,0)* versus annealing time are presented in Fig. 6. The dashed lines show the best-fits in terms of the JMAK-like equation yielding 11.8 at.% and 12.7 at.% for the strained and the strain-free samples, respectively. These values, at the first glance, compare remarkably well to those determined from *<B>=f(x)*. One has, however, to remember that the actual distribution of Cr atoms in the studied samples is not random. This can be already seen in Fig. 6 viz. for the non-annealed samples *x(0,0)* ≈13.5 at.% whereas the true concentration is 15.15 at.%. On the other hand, the data shown in Fig. 6 again show the difference between the two studied samples.

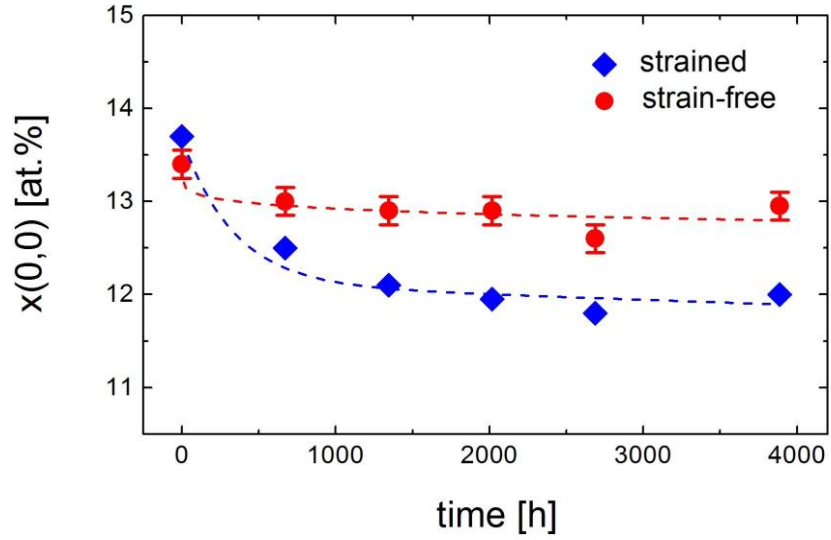

Fig. 6 Local Cr concentration, $x(0,0)$, vs. the annealing time. The lines represent the best fits in terms of the JMAK-like equation.

Another way of expressing the local concentration of Cr, $x_k$, is as follows:

$$x_k(at.\%) = \frac{<n_k>}{M} 100 \qquad (2)$$

where $M=8, 6, 14$ for $k=1, 2, 12$, respectively. This measure of the local concentration seems to be more adequate than the one outlined above because it takes into account all atomic configurations included in the spectra analysis except the (0,0) one. The $x_k$ – values obtained based on eq. (2) are displayed in Fig. 7.

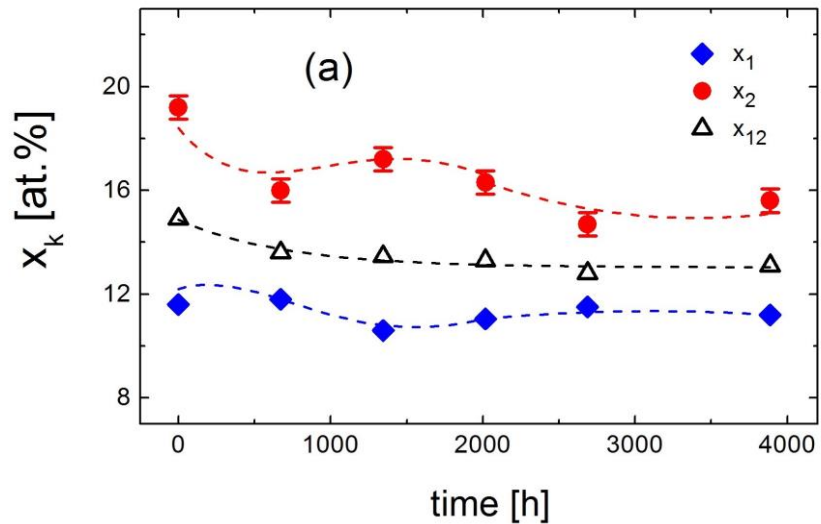

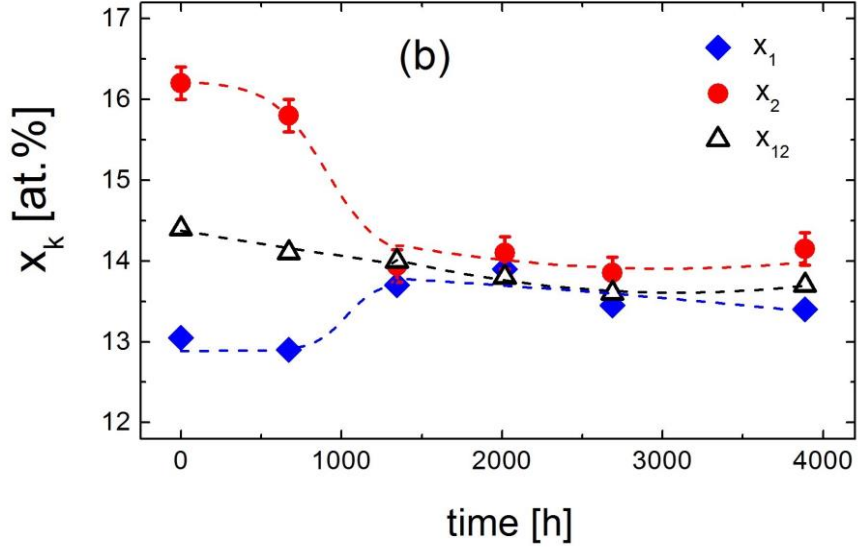

Fig. 7 Local concentrations of Cr as found based on equ. (2) vs. the annealing time for the strained (a) and the strain-free (b) samples. The lines are the guide to the eyes.

The behaviors seen in Fig. 7 reflect those found for the SRO-parameters which is expected from the definition. Nevertheless, it is interesting to observe the effect of annealing on the Cr concentration in both neighbor shells. One can clearly see similarities and differences between the two samples. Concerning the former, the average local concentration decreases with the annealing time which means that the number of Cr atoms within the 1NN-2NN neighborhood of the probe Fe atoms decreases. This effect can be understood in terms of the Cr atom clustering leading to a formation of the Cr-rich phase (α'). However, the local concentrations in the particular shells show a different behavior for the two samples. Namely, in the strained sample, the concentration of Cr in 1NN, $x_1$, is always lower than the average value, $x_{12}$, whereas the one in 2NN is remains always higher. A deviation in both cases amounts to ~1 at.%. For the strain-free sample a similar deviation from the average value is observed only in the start-up sample and in the one annealed for 672 h. In the samples annealed longer $x_1 \approx x_2$.

### 4.4. Magnetic texture

The analysis of the spectra with the procedure II gave also an information on the magnetic texture i.e. an average orientation of the magnetization vector relative to the direction of the gamma rays (normal to the sample's surface), $\theta$. This information is encoded in the ratio between the amplitude of the second and the third lines of a sextet, *C2/C3*. The knowledge of the latter enables determination of an angle $\theta$ from the following formula:

$$C2/C3 = \frac{4\sin^4\theta}{1+\cos^2\theta} \qquad (3)$$

The $\theta$ - values derived in this way are displayed in Tables 1 and 2. Evidently there is a meaningful difference between the strained (average value 52.8°) and the strain-free (average value 57.3°) samples, but in both cases $\theta$ does not depend on the annealing time.

## 5. Conclusions

It is evident from the present study that the phase decomposition took place in both samples, yet significant differences were revealed i.e. a metallurgical state of an Fe-Cr alloy has to be taken into account in order to properly describe and understand underlying mechanism(s). Based on the results revealed in this study the following conclusions can be drawn:

(1) The kinetics of the phase decomposition determined from the average hyperfine field is in line with the Johnson-Mehl-Avrami-Kolgomorov equation.

(2) The values of the Avrami exponent for the studied samples are the same (2.5), hence the mechanism of the phase decomposition, mixed nucleation and volume diffusion controlled growth, seems to be independent of the strain.

(3) The rate constant and the activation energy significantly depend on the strain: the latter is equal to 1666 and 1388 kJ/mol for the strained and the strain-free samples, respectively.

(4) Solubility of Cr in iron depends on the strain: its limit at 402°C was found to be 11.6 at. % for the strained sample and 13.2 at.% for the strain-free one.

(5) Essential differences in the studied samples were revealed in the behavior of the short-range parameters and/or local concentrations of Cr.

(5) Magnetic texture does not depend on the annealing time but it is different in the two samples.

### Acknowledgements

This work has been carried out within the framework of the EUROfusion Consortium and has received funding from the Euratom research and training programme 2014-2018 under grant agreement No 633053. The views and opinions expressed herein do not necessarily reflect those of the European Commission. This work was also supported by the Ministry of Science and Higher Education (MNiSW), Warsaw and the AGH University of Science and Technology, Krakow, Poland.